# Evidence for Supercurrent Connectivity in Conglomerate Particles in NdFeAsO$_{1-\delta}$

J D Moore, K Morrison, K A Yates,
A D Caplin, Y Yeshurun*, L F Cohen

*Blackett Laboratory, Imperial College, SW7 2AZ London*

J M Perkins, C M McGilvery, D W McComb

*Materials Department, Imperial College, SW7 2AZ London*

Z A Ren, J Yang, W Lu, X L Dong, Z X Zhao

*National Laboratory for Superconductivity, Institute of Physics and Beijing National Laboratory for Condensed Matter Physics, Chinese Academy of Sciences, PO Box 603, Beijing 100190, PR China*

**Abstract**

Here we use global and local magnetometry and Hall probe imaging to investigate the electromagnetic connectivity of the superconducting current path in the oxygen-deficient fluorine-free Nd-based oxypnictides. High resolution transmission electron microscopy and scanning electron microscopy show strongly-layered crystallites, evidence for a ~ 5nm amorphous oxide around individual particles, and second phase neodymium oxide which may be responsible for the large paramagnetic background at high field and at high temperatures. From global magnetometry and electrical transport measurements it is clear that there is a small supercurrent flowing on macroscopic sample dimensions (mm), with a lower bound for the average (over this length scale) critical current density of the order of $10^3$ A/cm$^2$. From magnetometry of powder samples and local Hall probe imaging of a single large conglomerate particle ~120 microns it is clear that on smaller scales, there is better current connectivity with a critical current density of the order of $5 \times 10^4$ A/cm$^2$. We find enhanced flux creep around the second peak anomaly in the magnetisation curve and an irreversibility line significantly below $H_{c2}$(T) as determined by ac calorimetry.

**Introduction**

The generic composition of the quaternary oxypnictides is LPMPn (L = La, Pr, Ce, Sm); M= Mn, Fe, Co, Ni; Pn = P, As. The crystal structure of the parent compound LaFeAsO is well-established, and the band structure, the multi-sheet Fermi surface of the doped (superconducting) compound, the density of states, the electron-phonon coupling, have all been calculated theoretically.[1,2,3,4,5] These are layered materials (and hence significantly anisotropic) in which in the undoped parent compound the (conducting) $Fe_2As_2$ layer is sandwiched between (insulating) $La_2O_2$ layers. Doping the $La_2O_2$ layer by replacing the $O^{2-}$ with $F^-$ provides extra positive charge in the insulating layer and negative charge in the conduction layer. The material is then a low carrier density semi-metal but with a high density of states near the Fermi energy (because of a van Hove singularity), and is on the verge of itinerant magnetism.[2] There is no theoretical consensus yet concerning the symmetry of the order parameter[3,6,7] nor whether the multiple sheets of the Fermi surface manifest themselves as a multi-gap superconductor. For the $LaFeAsO_{1-x}F_x$ compound, strong electron-electron correlations are suggested from the temperature-dependent resistivity, Seebeck coefficient, and thermal conductivity data.[8] Combined neutron scattering data, specific heat measurements and Hall data show that the carriers are predominantly electron-like, there are strong electron–electron correlations, and there is clear evidence for the formation of a spin density wave (SDW) gap in the undoped compound.[9] An upper limit for the electron carrier concentration of $1 \times 10^{21}$ $cm^{-3}$ is inferred from Hall data just above $T_c$. Encouraging for potential applications is that in this $T_c = 28.2$ K compound, the resistive transition in field suggests $H_{c2}$ is remarkably high, ~30 T. The coherence lengths[8] are short, ~35 Å, but not as short as in high temperature superconductors so intergrain connectivity may not necessarily be disrupted.

Recently a new family of oxypnictides have been prepared by high pressure synthesis $ReFeAsO_{1-\delta}$ without fluorine doping.[10] For Re= Nd, a dome-shaped phase diagram was found for $T_c$ versus oxygen deficiency. Here we study the magnetic properties of the compound from that series with close to the highest $T_c$ reported, $NdFeAsO_{0.85}$. The tunability of the superconductivity by oxygen deficiency in these materials strongly resembles the HTS oxide systems, and, as we show below, the electromagnetic connectivity of these materials also resembles the HTS case.

**Experimental Methods and Structural Characterisation**

NdFeAsO$_{0.85}$ was made by high pressure synthesis as described elsewhere.[11] In this paper we study a sintered bar of the material of dimensions 2.9mm x 1.7mm x 1mm which was about 85% dense, weighing 35mg and also we flake off powdered material from the bar. The particles flaked off the bar are a range of powder conglomerates, the largest being of the order of 100 microns across. Such particles are polycrystalline (the grain size is of order microns), but the crystallites within them are likely to better-bonded mechanically, and so perhaps also electrically better-connected, than the weakest bonding within the sintered bar. The powder conglomerates and smaller powder particles are all rather plate-like. Scanning electron microscopy (SEM – LEO 1525) was used to analyse the morphology of the sintered body. The SEM sample was produced by scraping the sintered body with the resulting powder being affixed to a stub using carbon tape. SEM images revealed that the sample does not consist of a single phase and energy dispersive x-ray (EDX) analysis was used to identify the chemical composition of particles with different morphologies. The superconducting phase was generally found to adopt platelet morphology (Fig. 1a). The EDX results suggest that crystallites of the superconducting phase are compositionally homogeneous, at least within the limits of SEM-EDX analysis. Particles exhibiting a partially delaminated layer structure were identified as being neodymium oxide (Fig. 1b). These type of particles are likely to be the source of the large paramagnetic background seen on the magnetisation curves. Transmission electron microscopy (TEM) images were obtained using an FEI Titan operated 300 kV. The powder scraped from the sintered bar was dropped on to a holey carbon coated copper grid. In the TEM, the majority of the electron transparent samples exhibited the platelet morphology observed in the SEM. The platelets appear to consist of well oriented layers and many have an amorphous phase that is 5-15nm wide surrounding the particles (Fig. 1c-d).

Electrical Transport measurements were made on the solid bar with dimensions 0.98mm x 1.7mm x 3.56mm using a conventional four probe technique. An Oxford Instruments vibrating sample magnetometer with an 8T superconducting magnet was used for magnetometry. Local Hall probe imaging was conducted using a scanning 5 micron InSb Hall cross micromagnetometer[12] and microcalorimetry was performed using a Si nitride membrane calorimeter as described in detail elsewhere.[13]

**Results and Discussion**

Electrical transport data (see lower inset to Fig. 2) shows that there are well connected current paths through the whole bar. The main figure 2 shows that the irreversible magnetisation Δm taken on the bar sample is larger than on the powder conglomerates at 25K. Contributions to Δm from circulating currents depend on the length scale on which they flow;[14] there may be full connectivity, with currents circulating around the entire sample, or smaller regions of strong connectivity, or (as is often the case in HTS materials) poor intergranular connectivity, leaving only the intragranular currents to contribute. If we take the bar dimensions as the length scale for the global magnetometry, we can estimate a lower bound (if the currents flow, at least in part, on a smaller scale, the estimate for $J_C$ is correspondingly higher) of $J_C = 10^3$ A/cm$^2$.

Figure 3 shows the results of Hall probe imaging a single 120 micron conglomerate. The Hall images are approximately 200 μm across and reveal the magnetic intensity at a constant height 20 μm above the sample. The lower inset shows the magnetic image of the fragment at 15K at 20mT. In the images shown here, black represents zero local magnetisation and red (yellow) represents regions of positive (negative) local magnetisation. At this temperature and field, the flux has a regular distribution, showing brightest yellow (largest magnetisation) in the centre of the image. This is shown most clearly by a line-scan of the field image, which has a classic Bean profile (with some averaging because the sensor is ~20 microns above the 120 micron size fragment) with partial penetration of flux at low field. The fragment only partially screens the field as would be expected from the geometry. Interestingly the upper inset in Fig. 3 shows the same fragment at 15K at 0.5T where the magnetic response is dominated by a large paramagnetic background (upper inset to Fig. 2 shows the *M-H* loop and this large background signal is clear). Indeed the upper image shown in Fig. 3 shows a well-defined paramagnetic region particularly on the right hand side of the sample. The profiles taken across the sample at 140mT and above confirm the presence of the paramagnetic signal. Note that the paramagnetic material is distributed inhomogeneously, and is likely to be a second phase such as the Nd oxide imaged in figure 1(b). Very similar paramagnetic signal has recently been interpreted as an intrinsic property of Nd based oxypnictides due to the existence of the magnetic Nd ions.[15] We believe that the inhomogeneity of the signal we see across the fragment rules this scenario out for the composition studied here.

From Figs. 2 and 3, using the Bean profile at full penetration (20mT) and the magnetic signal above the single fragment (in the far field magnetic dipole limit) we can approximately estimate $J_c$ in the single fragment to be of the order of 5 x $10^4$ A/$cm^2$ at 15K Again, this estimate is a lower bound, but probably a good one for the intergranular current density; there may well be a higher intragranular current density.

Figure 4 main figure plots the irreversibility line extracted from the irreversible magnetisation of the bar sample (see the top-right Fig. 4) using a $10^{-3}$ emu threshold criterion. Note that the irreversible magnetisation has a temperature-dependent, but field-independent, magnetic background which we subtract before estimating the $H_{irr}$ threshold. To extract $H_{c2}(T)$ we employed a microcalorimetry technique studying the same fragment imaged in Fig. 3. The heat capacity of the fragment was measured as a function of temperature in 0, 2, 4, 6 and 8T as the system was cooled and warmed. To highlight the contribution due to the superconducting transition, the normal background heat capacity was estimated as a polynomial fit to the data and subtracted. The resultant peaks observed (inset to main in Fig. 4) were used to determine the $H_{c2}(T)$ curve plotted in the main figure.[16] To reduce error due to thermal gradient across the sample/gauge an average value for $H_{c2}$ using the data taken when the sample was being cooled and warmed. Figure 4 also shows the position of the second magnetisation peak $H_{peak}$ determined as the maximum feature in d$M$/d$H$ at each temperature. The $H$-$T$ diagram strongly resembles that of many of the highly anisotropic high temperature superconducting systems.[17]

We use the method known as "sweep creep"[18] as first proposed by Pust *et al.*,[19] to determine the flux creep properties in this sample. We define the normalised dynamic creep rate S, as S = dlnΔM/dlnH' where H' is the sweep rate of the magnetic field, H'= dH/dt. By definition, S averages over two legs of the hysteresis loop, but permits rapid determination of the creep rate as a function of field. The bottom-right plot in Fig. 4 shows S and its variation with applied magnetic field in the bar at 35K. There is clearly an approximate inverse relationship between the variation of S(B) and Δm(B).  The first report of such a "mirror image" relationship between the creep rate S and ΔM was made in high temperature superconductors in 1992.[20] The flux creep observed in these oxypnictide samples is large, similar in magnitude to HTS materials and the strong enhanced creep rate associated with the second peak feature is also very pronounced in this system. A detailed study of vortex regimes in the oxypnictide superconductors based on dynamic creep rate[21] is beyond the scope of this present paper, but it is already clear that many of the issues as well as much of the interesting science are similar to those that arose with the cuprate materials.

**Conclusions**

Here we report on the connectivity issue in the new superconducting oxypnictide materials. We find that the macroscopic bar of the material appears to carry a finite $J_c$ which is a positive for possible applications of the material. Fragments taken from the bar show $J_c$ values at least an order of magnitude higher than this and from TEM and SEM studies it seems that these fragments are conglomerates of several small crystallites. Consequently intragrain $J_c$ may indeed be of the order of $10^5$ A/cm$^2$ which is encouraging although there are clearly connectivity issues that need to be surmounted. The vortex phenomenology strongly resembles that of the cuprate superconductors showing second peak features, enhanced creep rates and an irreversibility line well below the upper critical field.

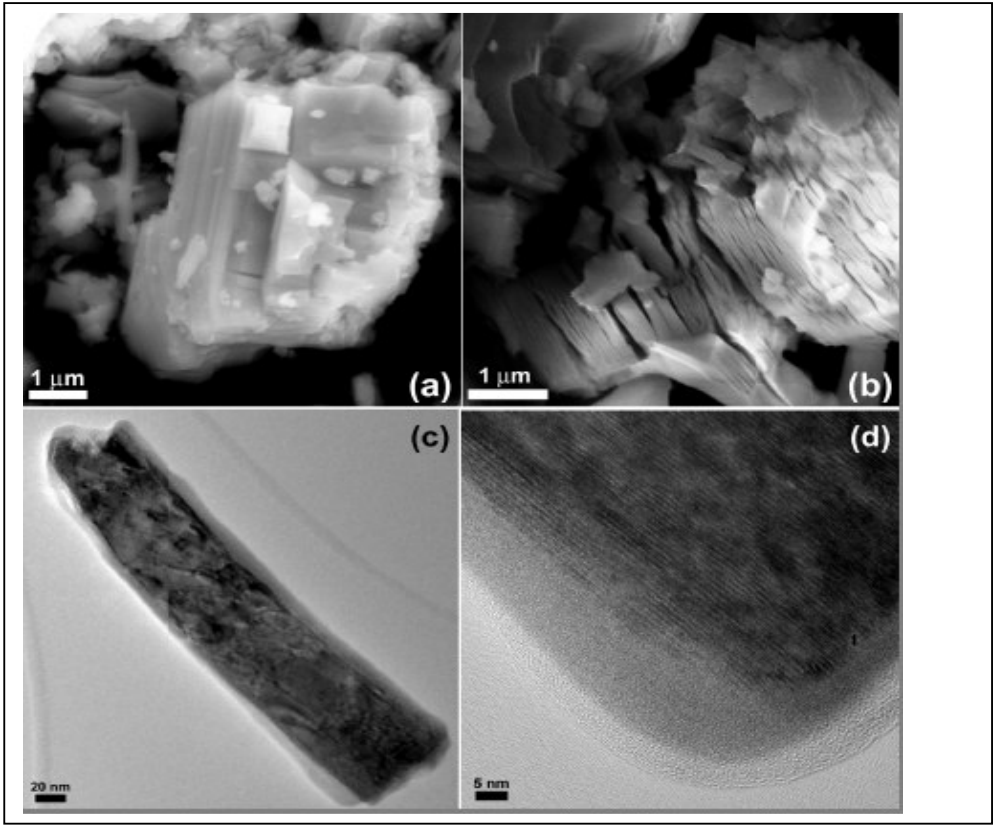

**Figure 1** a) FE-SEM image of typical NdFeAsO$_{1-\delta}$ particle from the as-sintered body. b) neodymium oxide particle, chemistry confirmed using EDX from various positions on the particle. This grain shows a different morphology to the more platelet-like superconducting grains. Furthermore, the fracture paths of the platelet NdFeAsO$_{1-\delta}$ grains show a possible layered structure. c) and d) low and high magnification TEM images of a single NdFeAsO$_{1-\delta}$ grain showing both evidence of a layered structure as well as an amorphous layer surrounding the particles.

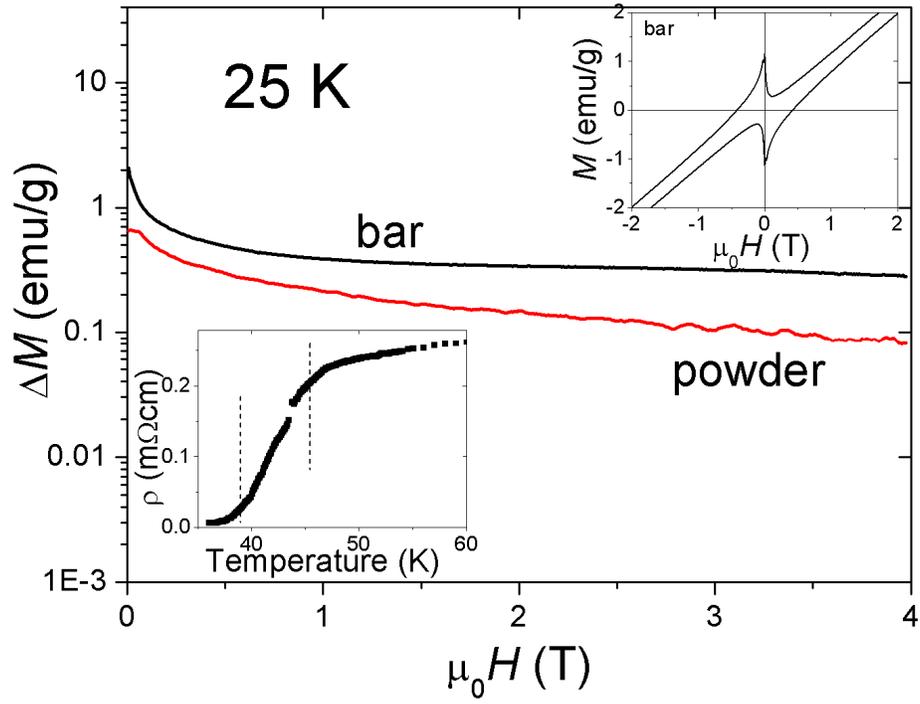

**Fig 2 –** Irreversible magnetisation versus magnetic field at 25K on the powder and the bar sample. Top inset shows the bar *M-H* loop at 25K with the paramagnetic background. Bottom inset shows resistivity versus temperature. The dashed lines mark the 10% and 90% points.

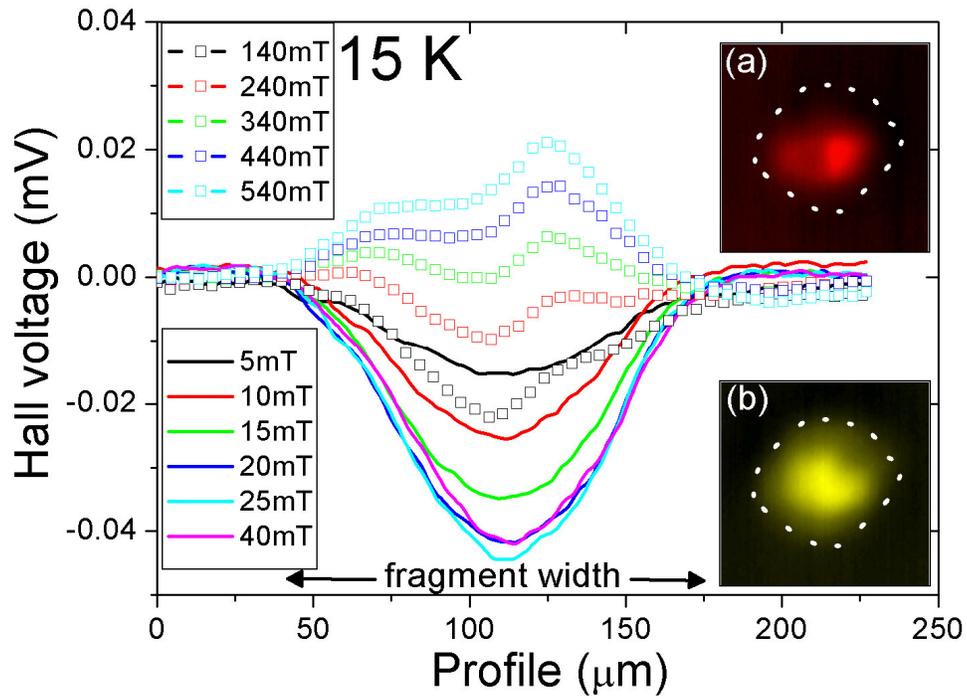

**Fig 3** The cross-sections (B-$\mu_0$H) taken from the Hall images at various applied magnetic fields on the 120 micron size fragment at 15K on the virgin field-leg. Inset (a) shows the full image at 0.54T (red is positive *M*) and inset (b) shows the full image at 20mT (yellow is negative *M*). The dotted lines indicate the sample border.

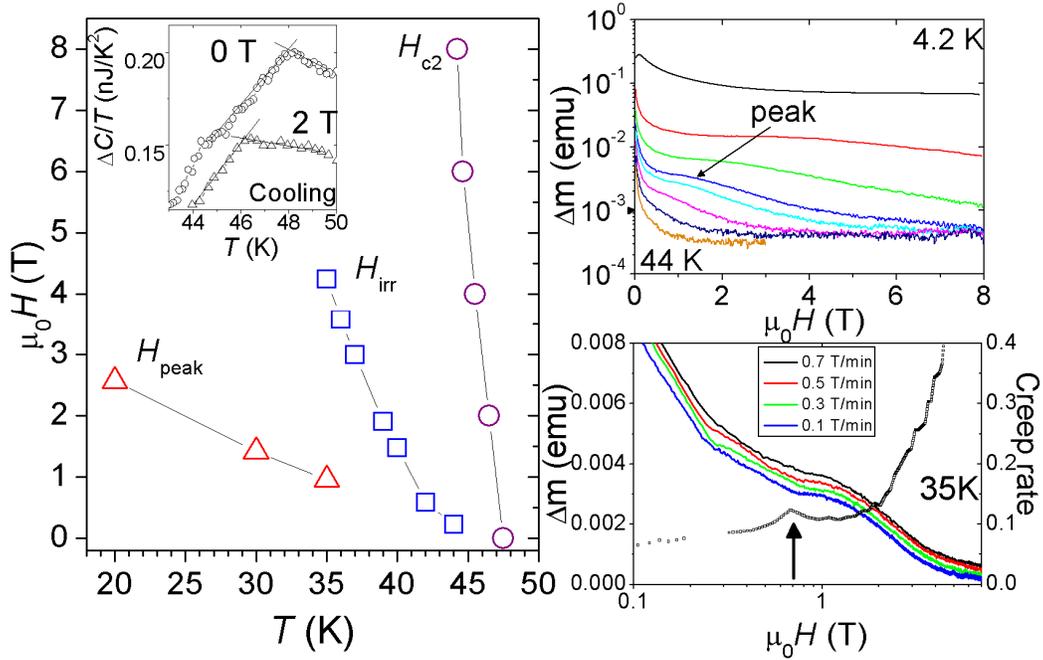

**Fig 4** (main) *H-T* diagram showing the irreversibility field $H_{irr}$(T) defined as the field at which the magnetic moment of the bar sample (32mg) falls below $10^{-3}$ emu, the peak field $H_{peak}$ versus temperature taken from the irreversible magnetisation loops of the bar sample, and the upper critical field $H_{c2}$(T) taken from ac calorimetry measurements of the fragment. The inset to the main figure shows examples of the isofield measurements of heat capacity after subtraction of background (non-superconducting contribution) where $\Delta C$ is defined as ($C_{total}$ - $C_{background}$). Top right figure shows $\Delta m$(H) (taken from the bar sample) for a range of temperatures (from top to bottom) 4.2, 20, 30, 35, 37, 40, 42 and 44K. Bottom right figure shows the normalised creep rate S (see text for definition) versus field at 35K taken from the bar sample (black curve) and (left axis) associated $\Delta m$(H) at 35K for the different sweep rates 0.1, 0.2, 0.3, 0.5 and 0.7 T/min.

* Permanent address: Institute of Superconductivity, Department of Physics, Bar-Ilan University, Ramat-Gan, Israel 52900